\documentclass[12pt]{iopart}

\usepackage[T1]{fontenc}
\expandafter\let\csname equation*\endcsname\relax 
\expandafter\let\csname endequation*\endcsname\relax
\usepackage{amsmath}
\usepackage{color}
\usepackage{graphicx}
\usepackage{hyperref}
\usepackage{xcolor}
\usepackage{cite}
\newcommand{\XXX}[1]{{\color{black} #1 }}
\newcommand{\lt}{\left }
\newcommand{\rt}{\right}
\newcommand{\dd}{\ensuremath{\mathrm{d}}}
\newcommand{\const}{\ensuremath{\mathrm{const.}}}
\newcommand{\refeq}[1]{(\ref{#1})}
\newcommand{\I}{\ensuremath{\mathrm{i}}}

\newcommand{\mH}{\ensuremath{\mathcal H}}
\newcommand{\mD}{\ensuremath{\mathcal D}}
\newcommand{\mC}{\ensuremath{\mathcal C}}
\newcommand{\mV}{\ensuremath{\mathcal V}}

\begin{document}

\title[Minimal model of diffusion with time changing Hurst exponent]{Minimal
model of diffusion with time changing Hurst exponent}

\author{Jakub {\'S}l\k{e}zak$^\dagger$ and Ralf Metzler$^{\ddagger,\sharp}$}

\address{$\dagger$ Hugo Steinhaus Center, Wroc{\l}aw University of Science
and Technology, Poland\\
$\ddagger$ Institute of Physics \& Astronomy, University of Potsdam, Germany\\
$\sharp$ Asia Pacific Centre for Theoretical Physics, Pohang 37673, Republic
of Korea}
\ead{jakub.slezak@pwr.edu.pl,rmetzler@uni-potsdam.de (Corresponding author:
Ralf Metzler)}

\begin{abstract}
We introduce the stochastic process of incremental multifractional Brownian
motion (IMFBM), which locally behaves like fractional Brownian motion with a
given local Hurst exponent and diffusivity. When these parameters change as
function of time the process responds to the evolution gradually: only new
increments are governed by the new parameters, while still retaining a
power-law dependence on the past of the process. We obtain the mean squared
displacement and correlations of IMFBM which are given by elementary
formulas. We also provide a comparison with simulations and introduce
estimation methods for IMFBM. This mathematically simple process is
useful in the description of anomalous diffusion dynamics in changing
environments, e.g., in viscoelastic systems, or when an actively moving
particle changes its degree of persistence or its mobility.
\end{abstract}

\section{Introduction and background}

The modern study of diffusive processes started at the beginning of the
20th century when Albert Einstein \cite{Einstein}, Marian Smoluchowski
\cite{Smoluchowski}, William Sutherland \cite{sutherland}, and Paul
Langevin \cite{LangevinEng} proposed the physical theory of Brownian motions
(later formalised by Norbert Wiener \cite{Wiener}), laying the foundations
for what is now the theory of stochastic processes and nonequilibrium
statistical physics \cite{brenig,landau,bennaim,livi}. Brownian motion is
characterised by a linear mean squared displacement (MSD) and a Gaussian
probability density function (PDF) \cite{vankampen,coffey}. While single
particle tracking already was well established in the early experiments of
Perrin \cite{perrin} and Nordlund \cite{nordlund}, with modern microscopic
techniques the stochastic motion of microscopic particles or even single
molecules can now be routinely recorded in complex environments such as
living biological cells \cite{Manzo2015}. Single trajectories are also
measured for moving cells, small organism, or even large animals, among many
other applications \cite{vilk,pccp,igorsoft,Manzo2015,pt,pt1, gorka}. Such
experiments demonstrate that in many complex systems the MSD is no longer
linear in time but follows the power-law form $\langle X^2(t)\rangle \propto
t^{2H}$, where the Hurst exponent $H$ distinguishes subdiffusion ($0< H<1/2$)
from superdiffusion ($1/2<H<1$) \cite{RWGuide}.

A widely used generalisation of the Einstein-Smoluchowski-Langevin theory of
Brownian motion is fractional Brownian motion (FBM) $B_H(t)$, the only
stochastic process which is Gaussian and exhibits power-law memory between its
increments\footnote{For $H\le1/2$ this expression has a strong singularity at
$t-s=w\to0$ which should be carefully accounted for. In integrals it is
interpreted as $\frac{\dd}{\dd w}(H|w|^{2H-1}\mathrm{sgn\ }w)$ \cite{Meerson},
similarly to equation \refeq{dIMFBMcov} below.}
\begin{equation}
\label{dFBMcov}
\left\langle\frac{\dd B_H(s)}{\dd s}\frac{\dd B_H(t)}{\dd t}\right\rangle=DC_H|t-s|^{2H-2},\quad C_H\equiv H(2H-1).
\end{equation}
Here the parameter $D$ is the (generalised) diffusion coefficient. Integrating
over this formula twice shows that the MSD has the power-law form $\langle B_H
(t)^2\rangle=Dt^{2H}$. In this context the coefficient $D$ can be interpreted
as the scale of the process for a given anomalous diffusion exponent $2H$. A
process closely related to FBM was originally proposed by Kolmogorov \cite{kolm},
and FBM was widely popularised by Beno\^it Mandelbrot and John van Ness in their
seminal paper \cite{MandelbrotNess,Graves}. Mandelbrot and van Ness were in fact
inspired not by diffusion, but Harold Hurst's studies of water flows \cite{Hurst},
economic cycles, and fractional $1/f^\alpha$ noises. In line with their
interdisciplinary approach, contemporary applications of FBM span very diverse
fields, from broadband network traffic \cite{MikoschNetworkTraffic} to the
structure of star clusters \cite{LomaxStarClusters}, and financial market
dynamics \cite{Gatheral}. Characteristics of subdiffusive FBM were, i.a.,
observed for the motion of submicron tracers in soft and bio matter
\cite{lene,lene1,weber,weiss,weiss1,jae_prl,glebprx}. Superdiffusive motion
consistent with FBM was observed in actively driven motion in biological
cells \cite{christine,christine1} and in movement ecology \cite{vilk,vilk1}.
We note that non-Gaussian forms of FBM measured, e.g., in biological cells
\cite{lampo,matthias1,ilpoprx}, may arise from FBM with randomly fluctuating
diffusion coefficient \cite{wei,wei1,agnes,nick}.

The cases of subdiffusion ($0<H<1/2$) and superdiffusion ($1/2<H<1$) have
differing physical interpretations and are only rarely observed together
\cite{Bouchaud,glebprx}. For $0<H<1/2$ the process exhibits negative memory
(see equation \refeq{dFBMcov}), a property referred to as antipersistence.
The negative dependence between increments and the covariance integrates to
zero, $\int_{-\infty}^\infty{\left\langle\frac{\dd B_H(s)}{\dd s}\frac{\dd B_H(s+t)}{\dd t}
\right\rangle}\dd t=0$. For $1/2<H<1$ the positive dependence between increments
causes the memory to be persistent, featuring a non-integrable tail of the
covariance function, such that $\int_1^\infty{\left\langle
\frac{\dd B_H(s)}{\dd s}\frac{\dd B_H(s+t)}{\dd t}\right\rangle}\dd t=\infty$ \cite{Beran}. It is worth adding that FBM is also used as a noise in generalised
Langevin equations \cite{Lutz,KouExp,jakub}.

Definition \refeq{dFBMcov}, while uniquely determining FBM and separating
it from other anomalous diffusion models, such as non-Gaussian subdiffusive
random walks \cite{RWGuide}, does not provide an explicit construction of the
FBM process. This can be achieved with one of the few equivalent integral
representations. The Fourier representation, equivalently for the increments
and the process itself, reads\footnote{Mathematically interpreted as the
stochastic integral over complex-valued white Gaussian noise $\dd Z(\omega)$.}
\begin{equation}
\label{FourierDef}
\dd B_H(t)=\frac{\sqrt{D}}{\gamma_H}\int\limits_{-\infty}^\infty\frac{\I\omega
\e^{\I\omega t}}{|\omega|^{H+1/2}}\dd t\ \dd Z(\omega)\quad\text{and}\quad B_H
(t)=\frac{\sqrt{D}}{\gamma_H}\int\limits_{-\infty}^\infty \frac{\e^{\I\omega t}
-1}{|\omega|^{H+1/2}}\dd Z(\omega)
\end{equation}
with the rescaling constant $\gamma_H\equiv\sqrt{{2\pi}}(\sin(\pi H)\Gamma(2H+1))
^{-1/2}$. This representation has the advantage of emphasising that FBM is a
model for a system at statistical equilibrium. Indeed, shifting time by some
$t_0$ only multiplies the integrand on the left by the complex phase $\exp(\I
t_0\omega)$ which leads to the same (real valued) PDF due to Gaussian
distribution isotropy of the $\dd Z$.
It also directly demonstrates that FBM is a $1/f^\alpha$
process with power spectral density $|\omega|^{1/2-H}$. Another representation
of FBM is the integral $\int_{-\infty}^t\big((t-s)^{H-1/2}_+-(-s)^{H-1/2}_+\big)
\dd B_s$ with $(t)_+\equiv\max(t,0)$; for more details see \cite{Picard}.

The strong symmetries of FBM make its Hurst index the unifying parameter
governing both its short and long time properties. This fact restricts some
of its applications---crucially for us it makes it impossible to describe an
increasing number of regime-switching anomalous diffusion systems in which
the anomalous diffusion exponent and the diffusivity change as functions of
time. Examples for such phenomena include the motion of a tracer in the
changing viscoelastic environment of cells during their cycle \cite{cycle}
or in viscoelastic solutions under pressure and/or concentration changes
\cite{barlow,clemens}, in actin gels with changing mesh size \cite{yael},
the motion of lipid molecules in cooling bilayer membranes \cite{jae_prl},
passive and active intracellular movement after treatment with chemicals
\cite{christine,seisen}, or intra- and inter-daily variations in the movement
dynamics of larger animals \cite{philipp}. Quite abrupt changes of $H$ and/or $D$ may
be effected by binding to larger objects or surfaces \cite{etoc,matthias1} or
multimerisation \cite{etoc,heller} of the tracer.

In order to overcome the limitations of FBM, multifractional Brownian
motion (MFBM) models were created, initially motivated by terrain modelling
\cite{PeltierMFBM}. A process is considered to be an MFBM if it resembles
FBM locally, i.e., its increments $\dd X(t)$ resemble increments of FBM
$\dd B_H$ with local parameters, a property called local self similarity
\cite{benassi1998identifying}. This definition does not specify any global
features of the process, such as the dependence between different $\dd X(s)$
and $\dd X(t)$---these can vary from model to model. Benassi, Roux and Jaffard
proposed an especially useful MFBM defined by substituting $H\to\mH_t$
into the right hand side of the integral in equation \refeq{FourierDef}
\cite{BenassiDef}. The simple mathematical form of the Fourier transform
allowed Ayache \etal to calculate the exact distribution of this model, a clear
advantage for practical applications; in particular it has MSD $Dt^{2\mH_t}$
\cite{Ayache2000cov}. For information about other MFBM variants see, e.g.,
the 2006 review by Stoev and Taqqu \cite{Stoev2006}. Recently, a Memory MFBM
(MMFBM) model was proposed in order to describe viscoelastic or persistent
anomalous diffusion with physically meaningful persistence of correlations
\cite{MMFBM}.  Another very closely related class of processes are linear time
series models with fractionally integrated noises (ARFIMA) and time dependent
coefficients \cite{Ray}. Inter alia MFBMs found applications in finance,
where it is natural to expect a time-dependence of the market dynamics
\cite{CorlayFinance,BianchiFinance,PengFinance,WuFinance}, and also network
traffic \cite{BianchiTraffic}, geometry of mountain ranges \cite{Echelard}
or atmospheric turbulence \cite{Lee13}, as well as heterogeneous diffusion
\cite{agnes}. Statistical methods for analysing MFBM models include wavelet
decomposition \cite{Wang01}, covariance and MSD analysis and testing
\cite{Balcerek20}, or neural networks \cite{Szarek22}.

In the standard MFBM models the history of the Hurst exponent $\mH_s,s<t$ does not
affect future observed displacements, only the local value of $\mH_t$ matters. This
feature does not affect the above-mentioned applications of MFBMs as they are
mostly concerned with the local roughness of the observed data, not its global
moments and correlations. This "amnesia" of the Hurst exponent is also the
reason behind MFBMs' simple mathematical structure. However, it is not a
desirable property for modelling diffusion in complex media where an evolving
$\mH_t$ should reflect the physical changes in the environment which determine
the further evolution of the process due to the inherent long-range memory. The
forgetfulness of MFBMs also forces the trajectories to "bend" to an evolving
$\mH_t$: Rapid changes of $\mH_t$ lead to rapid, jump-like changes of the
trajectory $X(t)$. In physical and biological contexts we would rather expect
that---in the same manner as for the diffusivity---changes of $H$ (even rapid)
should lead to a gradual response to a new environment due to the governing
long-range memory structure.

In the following we introduce and state the fundamental properties of a minimal
model for FBM with a time-evolving Hurst exponent in section \ref{sec2}. We then
present concrete results for a step-wise change of the Hurst exponent and the
diffusion coefficient in section \ref{sec3}. Finally we discuss our results in
a broader context in section \ref{sec4}. Simulation and estimation methods for the model are shown
in the Supplementary material.

\section{Definition of the minimal model}
\label{sec2}

In our approach to establish a minimal model to resolve the question of how
we can model anomalous diffusion of the FBM type with long-range correlations
and time-evolving transport coefficients $H$ and $D$ we want to preserve
the simple mathematical structure of the existing MFBM models but modify
this structure such that it does not directly affect the position of the
particle but reflects the gradual influence on the particle dynamics following
environmental changes as mediated by the memory structure of FBM. To this end
we consider an FBM-type diffusion for which the change in the environment
leads to changes of the Hurst exponent and diffusion coefficient. \XXX{ It is
then physically more natural to assume that this will not cause the whole
trajectory to "switch" to a
new $H$, but only affect the new increments after the change.} In different
words, changes of $H$ should lead to direct changes of the increments and
to only indirect changes of the position.

\begin{figure}
\centering
\includegraphics[width=16cm]{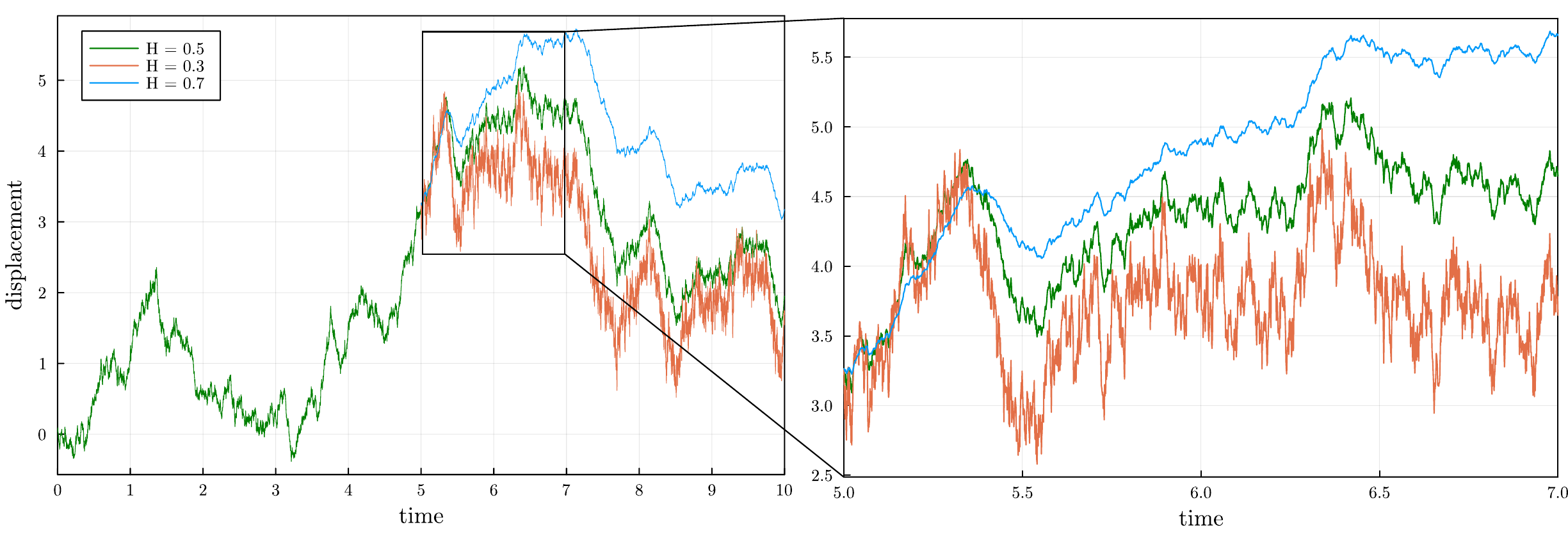}
\caption{Comparison of three trajectories, which represent normal Brownian
motion until $t=5$ and then switch to subdiffusion, superdiffusion, or
stay Brownian. The trajectories are based on the same realisation of the
stochastic noise. We see how a low $H$ ($<1/2$) introduces antipersistence by
amplifying the periods in which the trajectory turns back and, conversely,
how high $H$ ($>1/2$) introduces a persistence by amplifying excursions
in the same direction. The right panel shows a zoom into the part of the
trajectory right after the switching of $H$. \XXX{The simulations were performed using our introduced model with time changing $H$, the procedure is explained in the Supplementary Material.}}
\label{trajComp}
\end{figure}

The simplest way of fulfilling this requirement is to modify the memory
structure of FBM to
\begin{equation}
\label{dIMFBMcov}
\left\langle\frac{\dd B_\mH(s)}{\dd s}\frac{\dd B_\mH(t)}{\dd t}\right\rangle =\sqrt{\mD_s\mD_t}C_{\mH_s,\mH_t}|t-s|^{
\mH_s+\mH_t-2},
\end{equation}
where the rescaling constant $C_{\mH_s,\mH_t}$ ($C_{H,H}=C_H$) is to be
determined at the end of this section. We also require the process to
be Gaussian, as is FBM. Thus, the process \refeq{dIMFBMcov} is uniquely
determined, as there is only one Gaussian variable with a given covariance (and
zero mean). The resulting dynamic has a Hurst index defined by the arithmetic
mean $H\to(\mH_s+\mH_t)/2$ and a diffusion coefficient given by the geometric
mean $D\to\sqrt{\mD_s\mD_t}$. For any period with constant parameters $\mH_t=H$
and $\mD_t=D$ this process clearly reduces to a standard FBM, and thus this
process belongs to the broad class of MFBMs. As it is defined through its
increments we will call it incremental MFBM (IMFBM). We demonstrate how a
changing Hurst exponent affects IMFBM trajectories in figure \ref{trajComp}.
We note that the trajectory of MBFM would lead to a discontinuity at the
point where $H$ is changing.

Equation \refeq{dIMFBMcov} completely determines the memory structure of IMFBM.
The quantity $\langle\dd B_\mH(s)\dd B_\mH(t)\rangle$ can be interpreted as a
response function $r=r(s,t)$, that determines to which extent any past
infinitesimal change $\dd B_\mH(s)$ influences linearly the current change
$\dd B_\mH(t)$: the feedback is negative, i.e., $r<0$, for $\mH_s+\mH_t<1$ and
positive ($r>0$) for $\mH_s+\mH_t>1$. A heavy, non-integrable tail of $r$ is
present when $\mH_s+\mH_t>1$.

Assuming we start our observation at 0, $B_\mH(0)=0$, the MSD of IMFBM can be
obtained from the integral\footnote{This formulation is connected to Riemman
Liouville FBM as originally formulated by L{\'e}vy \cite{levy,MandelbrotNess}.}
\begin{equation}
\label{MSDint}
\langle B_\mH(t)^2\rangle=\lt\langle\lt(\int_0^t\dd B_\mH(s)\rt)^2\rt\rangle
=\int_0^t\int_0^t\langle\dd B_\mH(s_1)\dd B_\mH(s_2)\rangle,
\end{equation}
where the integrand is given by relation \refeq{dIMFBMcov}. \XXX{Analogously,
to obtain the full covariance $\langle B_\mH(s)B_\mH(t)\rangle$ only the limits
of the integrals are changed to $\int_0^s\int_0^t$}. Note
that the result does not depend on the past of $\mH_t,\mD_t,t<0$. If the
evolution of the system initiated at some $t_0<0$ and we started observing it
at time $t=0$ the measured displacements $B_\mH(t)-B_\mH(0)$ would be the same
as in \refeq{MSDint}. This is a practical feature based on the stationarity,
ingrained in \refeq{dIMFBMcov}, of the underlying displacements, due to which
IMFBM depends only on observed quantities.

From equation \refeq{MSDint} it is also apparent that in the special case
$\mH_t=H=\const$ and $\mD_t\neq\const$ the process reduces to
\begin{equation}
B_\mH(t)=\int_0^t\sqrt{\mD_s}\dd B_H(s),
\end{equation}
which is a natural choice of extending FBM to incorporate a time-varying
diffusivity \cite{MMFBM}. When $\mH_t=1/2$ this integral reduces to scaled
Brownian motion, a widely used Markovian model of diffusion with time-evolving
diffusivity \cite{Jeon14,lim}.

\begin{figure}
\centering
\includegraphics[width=7.6cm]{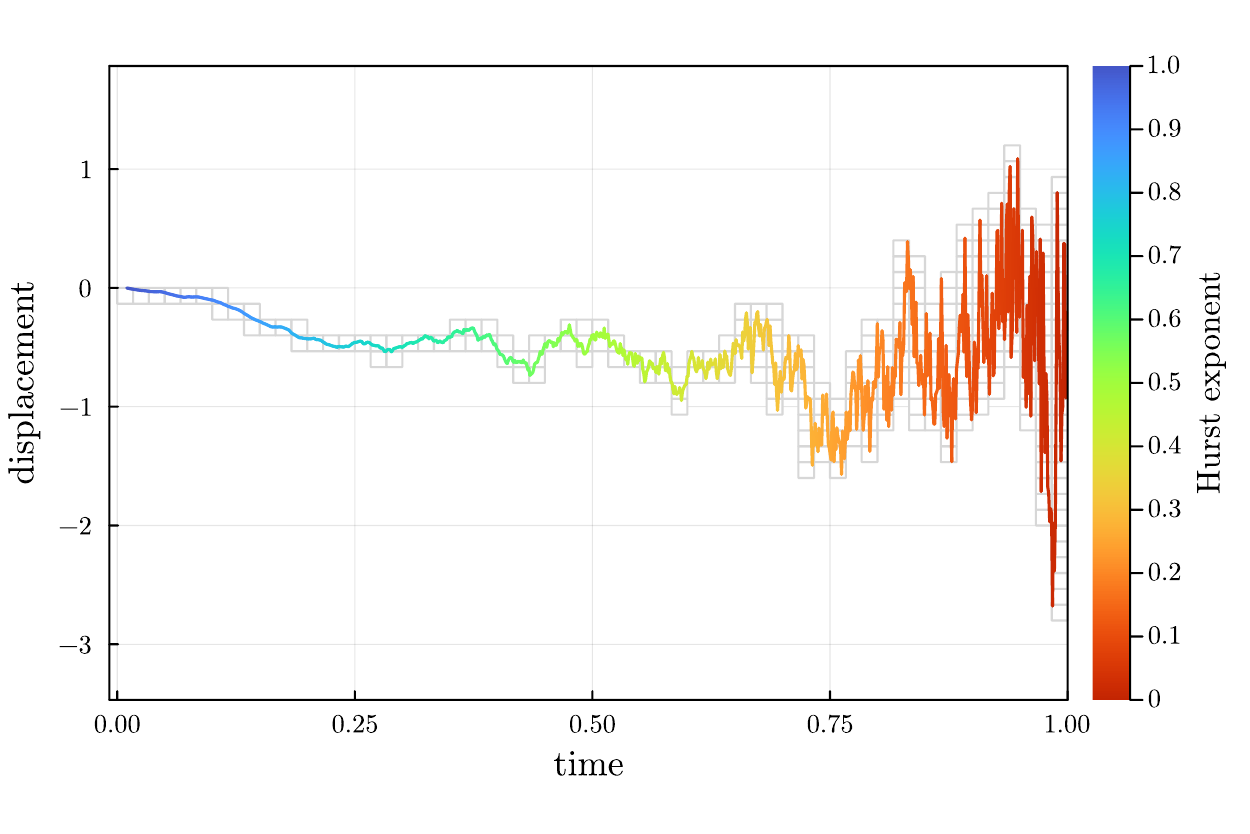}
\includegraphics[width=7.6cm]{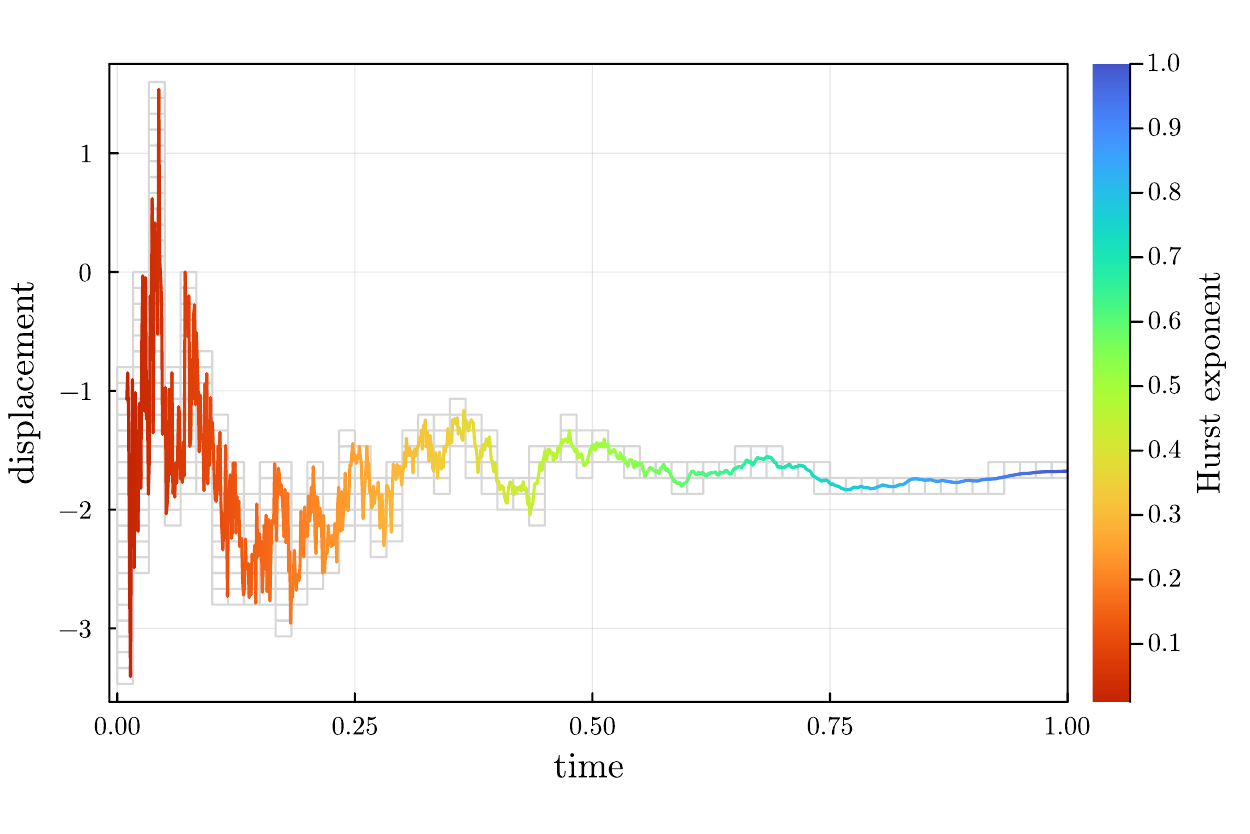}
\caption{Trajectory of IMFBM for which the Hurst exponent decreases linearly
from 1 to 0 (left panel) or increases linearly from 0 to 1 (right panel).
The grey areas show boxes with fixed size $\epsilon$, which cover the
trajectory. The more irregular the trajectory is locally, the more it
"covers" the space. In the limit $\epsilon\to 0$ the number of required
boxes locally increases proportionally to the fractal dimension,
$\epsilon^{-(2-\mH_t)}$ \cite{falconer}.}
\label{trajFrac}
\end{figure}

The above example shows that the choice of time changing diffusivity
$D\to\sqrt{\mD_s\mD_t}$ appears quite natural. The second part of the IMFBM
definition, the choice of $H\to(\mH_s+\mH_t)/2$ can be interpreted as imposing
a linear dependence on the Hurst exponent history. The current increment
$\dd B_\mH$ is correlated with a past increment with a weight proportional to
$|t-s|^{\mH_s}\dd s$. This relation is linear only for substitutions of the
form $H\to\lambda\mH_s+(1-\lambda)\mH_t$ with $0\le\lambda\le 1$. Giving equal
weight to past and present, $\lambda=1/2$, appears as a reasonable default
choice.
It also makes sure that trajectories of IMFBM (analogously to other MFBMs)
locally resemble FBM with parameters $\mD_t,\mH_t$, thus preserving the
local roughness of the trajectory: for smooth changes of $\mH$ and $\mD$
it has local fractal dimension $2-\mH_t$\footnote{It fulfils the conditions
for a fractal dimension given in \cite{BenassiFractDim}.}, as illustrated
in figure \ref{trajFrac}.

\begin{figure}
\centering
\includegraphics[width=5cm]{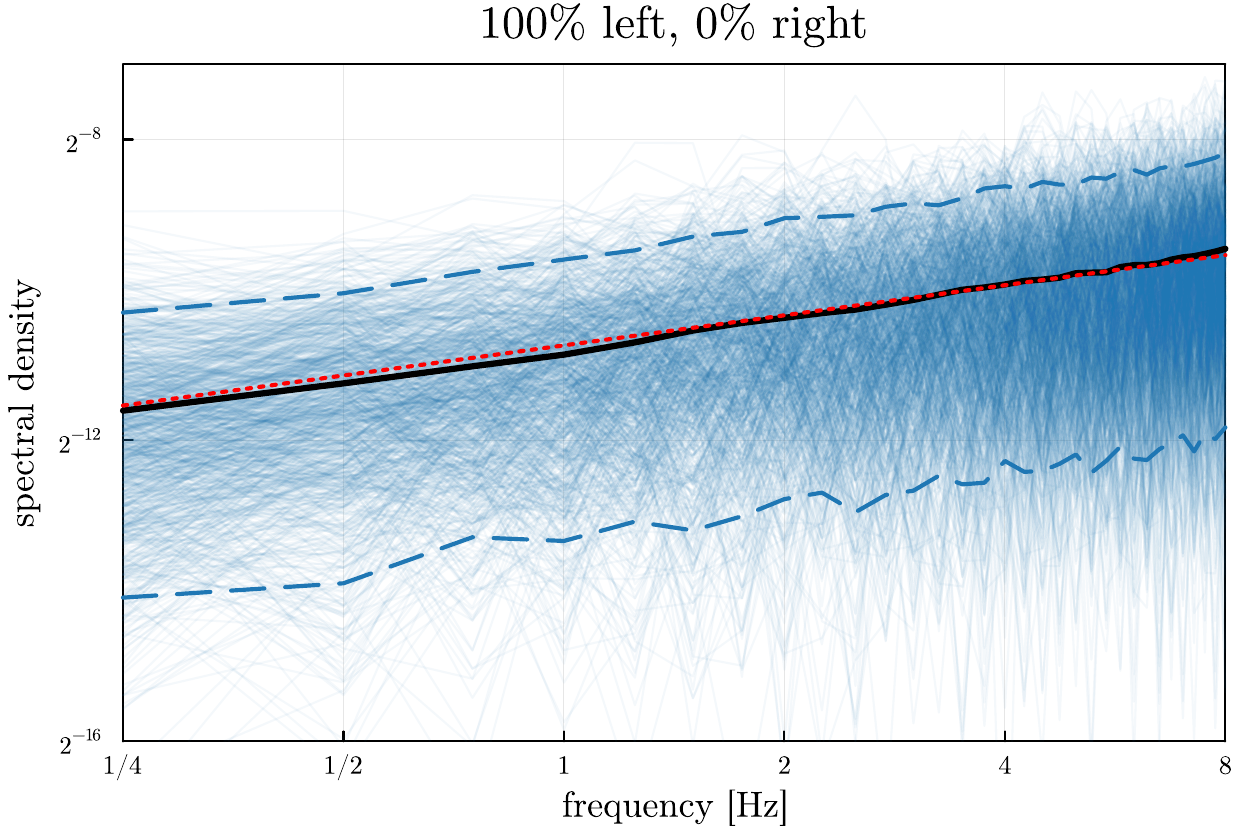}
\includegraphics[width=5cm]{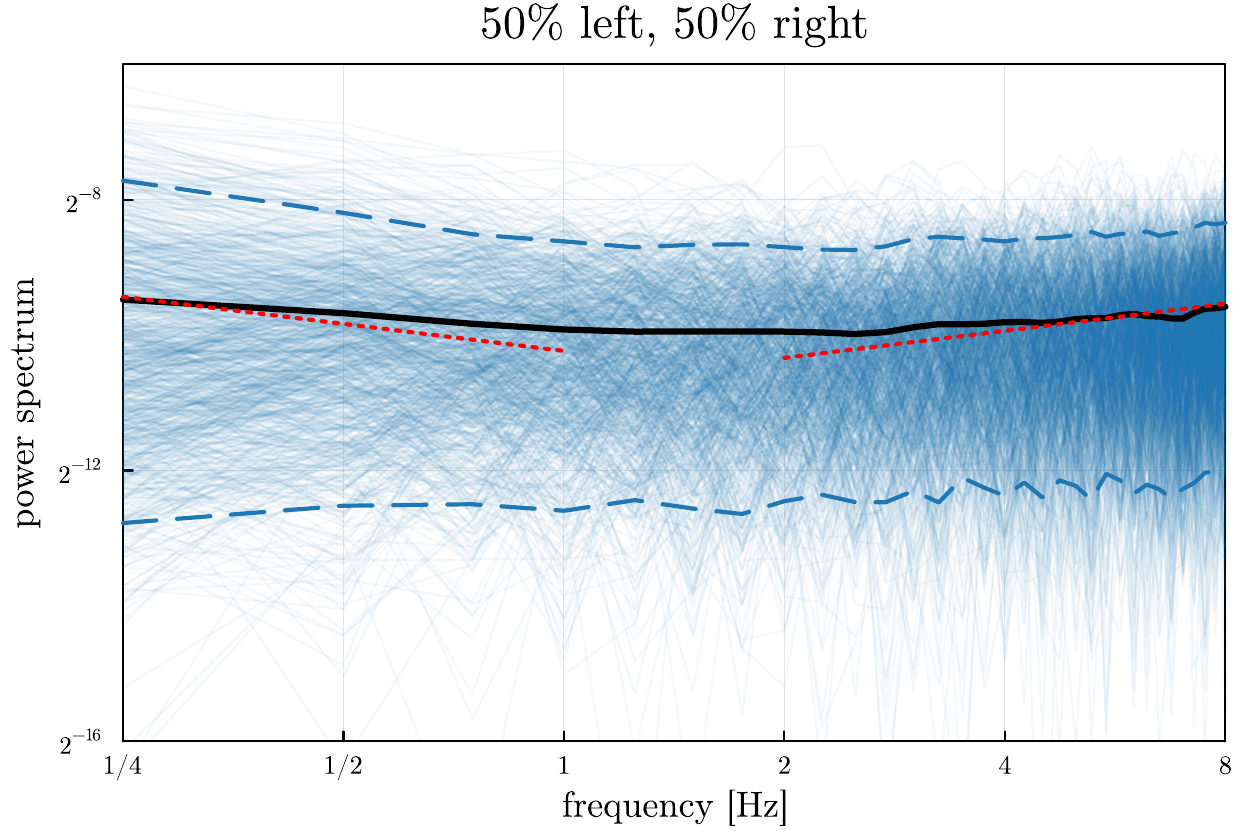}
\includegraphics[width=5cm]{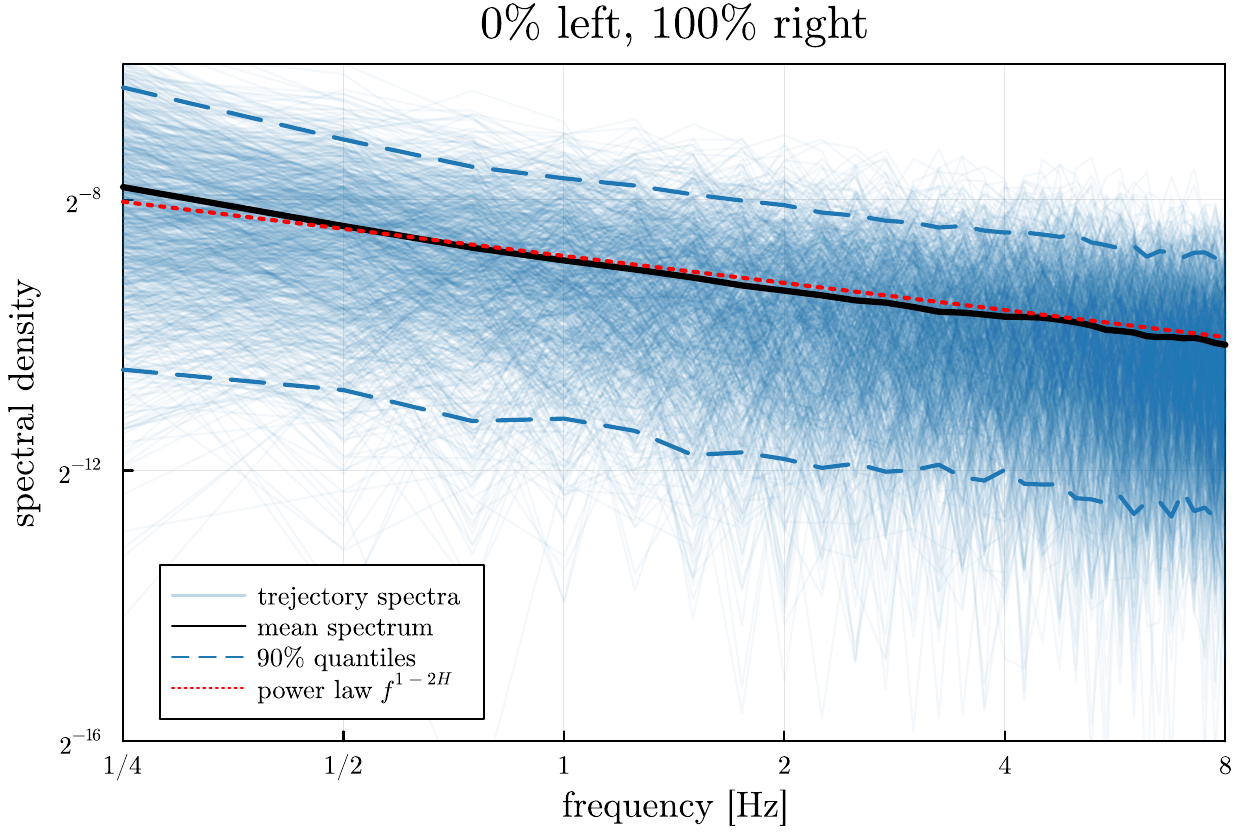}
\caption{Spectral densities calculated from a window of 1000 differentiated
IMFBM trajectory realisations with a single change of parameters from $H_1=
0.3,D_1=1$ to $H_2=0.7,D_2=16$ (see also section \ref{stepwiseChanges}). In
the Left panel the window is fully in the range of $H_1=0.3$ and the power law
spectrum agrees with the one of classical FBM. In the Right panel, analogously,
we see the same for a window fully in the range of $H_2=0.7$. In the Centre
panel we observe a mixed spectrum which on the left side is mostly affected
by the subdiffusive $H_1=0.3$ and on the right by the superdiffusive $H_2=0.7$.}
\label{spec}
\end{figure}

Returning to equation \refeq{dIMFBMcov}, we note that similar to
\refeq{dFBMcov} for FBM, it does not provide a representation (a direct
construction) of the process.  In fact, without such a representation, one
cannot even be sure that such a process exists mathematically. Due to its
elegant mathematical form the Fourier definition \refeq{FourierDef} of FBM
turns out to be useful here, as well. We simply substitute $D\to\mathcal
D_t$ and $H\to\mH_t$ into the left integral in \refeq{FourierDef} and
get\footnote{The process here is understood as a random generalised function,
i.e., this definition uniquely determines all integrals $\int \phi(t)\dd
B_\mH(t)$ for bounded $\phi$ with bounded support.}
\begin{equation}
\label{FourierRepr}
\dd B_\mH(t)\equiv\frac{\sqrt{\mD_t}}{\gamma_{\mH_t}}\int\limits_{-\infty}^
\infty\frac{\I\omega\e^{\I\omega t}}{|\omega|^{\mH_t+1/2}}\dd t\ \dd Z(\omega).
\end{equation}
By construction, this representation makes sure that the process is Gaussian,
as a combination of Gaussian increments $\dd Z(\omega)$. For $\mH_t,\mD_t\neq
\const$ the process is non-ergodic and non-stationary. Its increments belong
to the class of evolutionary spectra processes \cite{Priestley} with power-law
time dependent spectrum $\mathrm{sgn}(\omega)|\omega|^{1-2\mH_t}$, see figure
\ref{spec}. This power-law distribution of the probability mass over frequencies
determines the local fractal properties of the process, which are locally like
those of regular FBM.

It follows from the linearity of the Fourier transform that the covariance of
IFBM fits our requirement \refeq{dIMFBMcov}; essentially this process has the
two-point Fourier amplitude $\propto\sqrt{\mD_s\mD_t}|\omega|^{1-\mH_s-\mH_t}$
and it enforces \refeq{dIMFBMcov}. Using relation \refeq{FourierRepr} we can
determine the rescaling coefficient $C_{\mH_s,\mH_t}$: this can be achieved
using integral tables from which we find $C_{\mH_s,\mH_t}=c_{\mH_s,\mH_t}C_{
(\mH_s+\mH_t)/2}$ with $c_{\mH_s,\mH_t}\equiv\gamma_{(\mH_s+\mH_t)/2}^2/
(\gamma_{\mH_s}\gamma_{\mH_t})$. For periods of constant $H$, $c_{H,H}=1,
C_{H,H}=C_H$; otherwise the ratio $c$ weakens the dependence by a factor
increasing with the difference between $\mH$ values which accounts for the
varying local amplitudes of different trajectory segments.

\section{Stepwise changes of Hurst exponent and diffusion coefficient}
\label{stepwiseChanges}
\label{sec3}

The simplest---and essential---practical example for an evolving $H$ concerns
the switching between one distinct type of environment to another, and for
which we can neglect the influence of short transition periods. In the model
we then have that $\mH_t,\mD_t$ reduce to step functions with values $(H_1,
D_1)$, $(H_2,D_2)$, etc., at fixed intervals. As in any of the intervals with
constant $\mH_t,\mD_t$ the increments of IMFBM are---in isolation---equivalent
to FBM increments, the corresponding pieces of trajectory are identical to
segments of FBM with corresponding parameters $H_j,D_j$. If such a segment
were measured without the rest of the trajectory the \emph{increments\/} would
be indistinguishable from those of normal FBM. \XXX{Any statistical property,
based on ensemble-averages, time-averages, or both, will be the same.} When
multiple such segments are
experimentally observed, the IMFBM model provides additional information
about their codependence on the level of the \emph{position}. \XXX{Namely, any
two trajectory segments depend on each other through their compound Hurst
exponents and
diffusivities,} $(H_j+H_{j+1})/2$ and diffusivity $c_{H_j,H_{j+1}}\sqrt{D_jD_
{j+1}}$. Additionally, the full memory structure of a trajectory $B_\mH$ for
any number of transitions can be expressed by the direct formula obtained
from the covariance integral \refeq{MSDint}, see equation (S3).

To better understand the behaviour of this model let us consider a simple
concrete case crucial for many applications. A single transition between
two pairs of values $(H_i,D_i)$ at time $\tau$ corresponds to the protocol
\begin{equation}
\mH_t=\begin{cases}H_1, & t\le\tau,\\H_2, & t>\tau,\end{cases}\qquad
\mD_t=\begin{cases}D_1, & t\le\tau,\\D_2, & t>\tau.\end{cases}
\end{equation}
The associated MSD reads
\begin{equation}\label{stepMSD}
\langle B_\mH(t)^2\rangle=\begin{cases}D_1t^{2H_1}, & t\le\tau,\\[0.32cm]
D_2(t-\tau)^{2H_2}+D_1\tau^{2H_1} &\\
+c_{H_1,H_2}\sqrt{D_1D_2}\lt(t^{H_1+H_2}-(t-\tau)^{H_1+H_2}-\tau^{H_1+H_2}\rt),
& t>\tau.\end{cases}
\end{equation}
The cross term for $t>\tau$ has the asymptotic $\sim(H_1+H_2)\tau c_{H_1,H_2}
\sqrt{D_1D_2}t^{H_1+H_2-1}$. Thus, as expected, the MSD is dominated by $H_2$
at long times, $ \langle B_\mH(t)^2\rangle\sim D_2t^{2H_2}$, $t\to\infty$.
When $H_1+H_2<1$ the cross term disappears completely at long times. In the
opposite regime $H_1+H_2>1$ its remaining presence is indicative of the long
memory in the system.
Shortly after the transition at, $t=\tau+\delta$ with $\delta\to0$, the
asymptotic expansion of the MSD reads
\begin{equation}
\langle B_\mH(t)^2\rangle\sim D_1\tau^{2H_1}+D_2\delta^{2H_2}+(H_1+H_2)c_{H_1,
H_2}\sqrt{D_1D_2}\tau^{H_1+H_2-1}\delta-c_{H_1,H_2}\sqrt{D_1D_2}\delta^{H_1+H_2}.
\end{equation}
Among the three terms depending on $\delta$ the one with the smallest exponent
dominates at $\delta\to0$.

\begin{figure}
\centering
(a)\includegraphics[height=4.8cm]{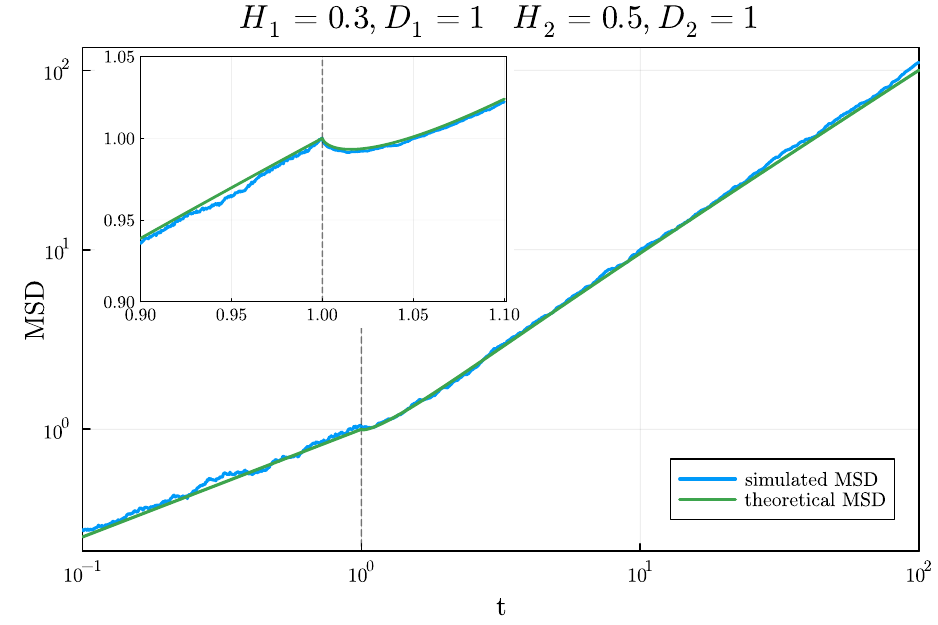}
(b)\includegraphics[height=4.8cm]{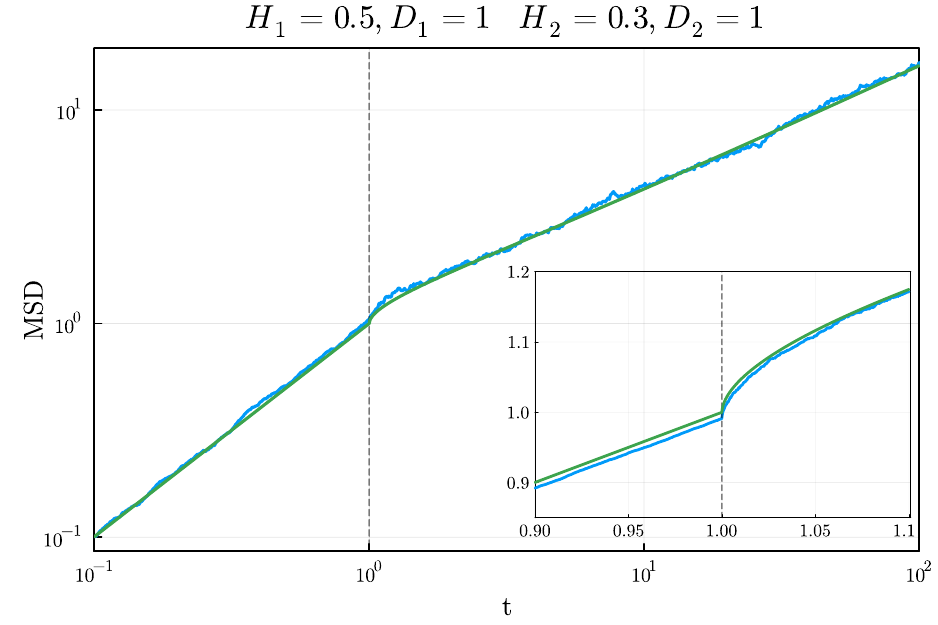}
(c)\includegraphics[height=4.8cm]{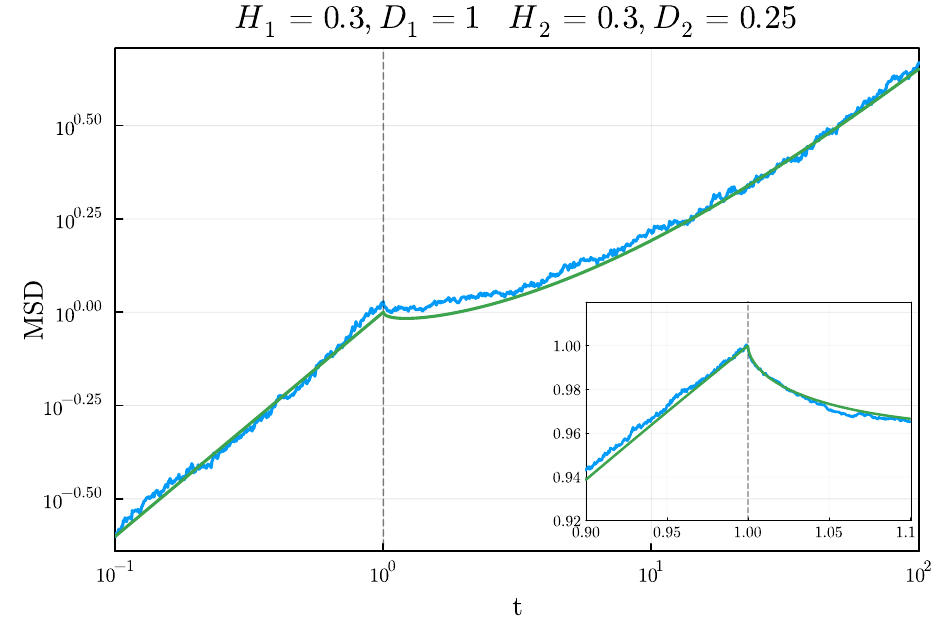}
(d)\includegraphics[height=4.8cm]{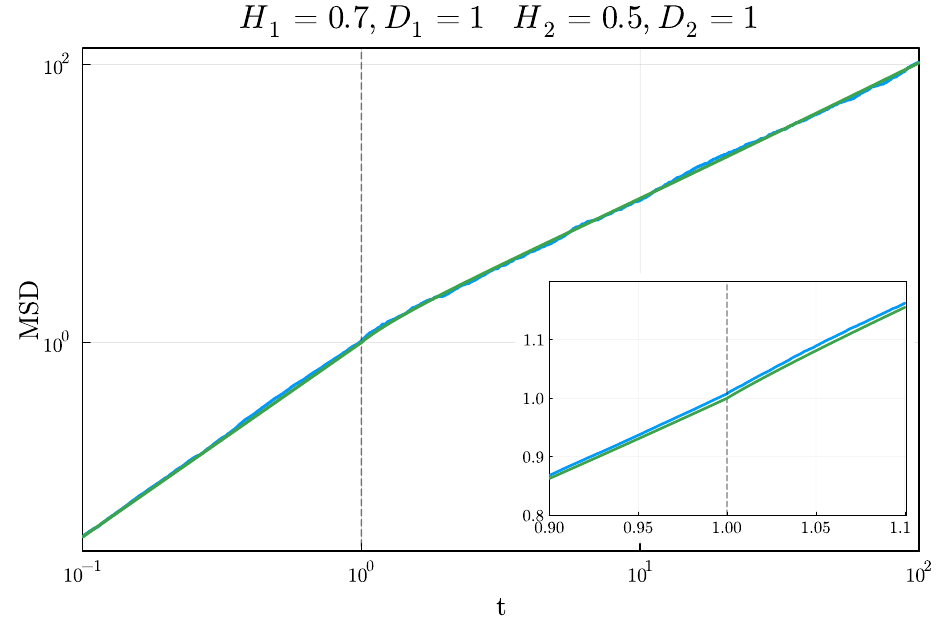}
(e)\includegraphics[height=4.8cm]{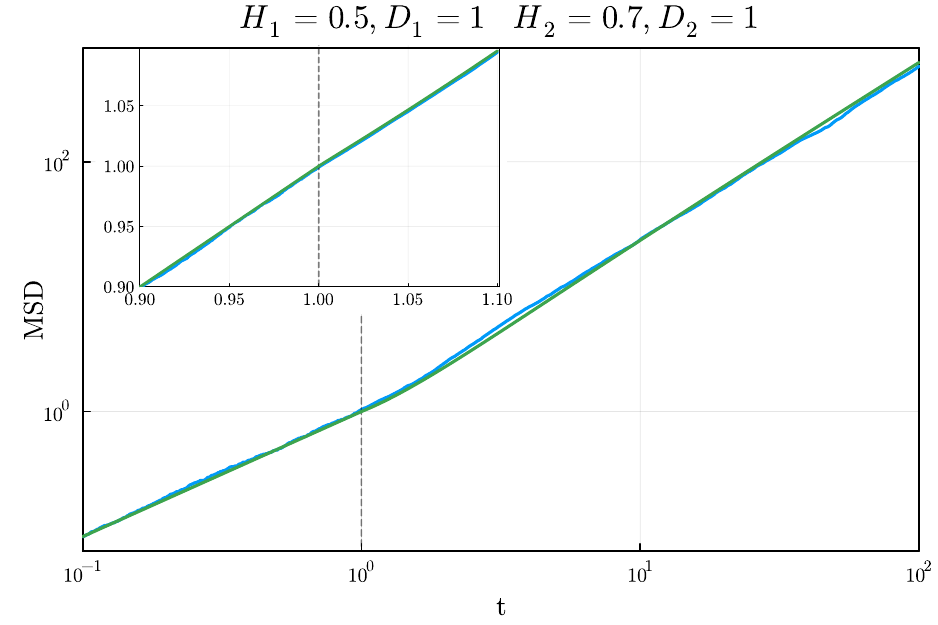}
(f)\includegraphics[height=4.8cm]{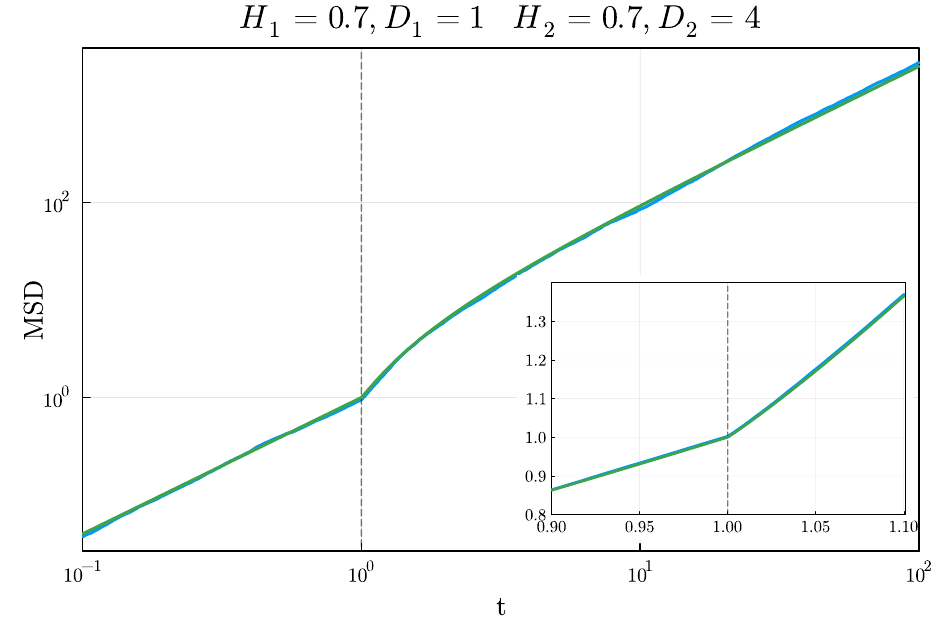}
\caption{Illustration of different types of transitions at $t=1$: \XXX{(a)
subdiffusion to normal diffusion, (b) normal to subdiffusion, (c) subdiffusion
to subdiffusion with different diffusivity, (d) superdiffusion to normal
diffusion, (e) normal diffusion to superdiffusion, and (f) superdiffusion to
superdiffusion with different diffusivity.}
The MSDs in the main plots were estimated
using $10^3$ trajectories. In order to visualise details of the motion,
in the inset plots with larger samples of $10^5$ were used.}
\label{MSDplot}
\end{figure}

Interestingly, in the case of weakening subdiffusion $H_1+H_2<1$, $H_1<H_2$, or
subdiffusion with decreasing diffusivity $H_1=H_2<1/2$, $D_2<D_1$, after the
transition at $\tau$ the MSD locally decreases. This may at first be seen as a
paradox, taking into account that displacements after time $\tau$ do increase,
$\langle(B_\mH(t+\tau)-B_\mH(\tau))^2\rangle=D_2t^{2H_2}$. However, a similar
behaviour can also be observed for other models with time-dependent
antipersistence and is caused by the fact that locally antipersistence dominates
persistence, $t^{2H_1}\ll t^{2H_2}$ for $H_2<H_1,t\to 0$. Thus, the tendency to
reverse the progression wins until a new piece of trajectory accumulates
sufficient weight. The same effect occurs for decreasing $D$ as then new
increments have less weight due to their smaller amplitudes. Different examples
of Hurst exponent transitions are shown in figure \ref{MSDplot}.

\section{Discussion}
\label{sec4}

The subject of anomalous diffusion is sometimes called a "jungle" of models
which alludes to the richness and variety of mathematical tools it offers
but also to common difficulties in deciphering the ones best suited to
a given system. New methods of describing transient diffusion phenomena
introduce additional time dependencies of the transport parameters which
adds yet another layer to the already complex modelling problem. This is
why it is crucial to have at disposal models as simple as possible, and
which preserve as much as possible from the elegant symmetries of anomalous
diffusion processes---this is what made them useful in the first place.

What is presented here is an attempt at introducing a time dependence
into FBM while keeping its memory structure as simple as possible. Such
processes---multifractional FBMs---have already been developed but they were
not widely used for describing diffusion phenomena and existing models do not
have probabilistic features which we would expect from those for physical
and biological systems. The introduced IMFBM models a particle diffusing
in a complex environment for which conditions change in time and after the
transition new displacements are governed by new diffusivity and new Hurst
exponent while also keeping the memory of its history before the
transition\XXX{; see table \ref{table} for a
comparison between classical MFBM models and IFBM.} It
has two central features: the geometric averaging of diffusivities and
arithmetic averaging of Hurst exponents (see \refeq{dIMFBMcov}) which are
fully experimentally verifiable and distinguish it from other MFBM models.

\begin{table}[]\centering
	\begin{tabular}{r|ll}
		\multicolumn{1}{l|}{} & classical MFBMs      & IMFBM                \\ \hline
		dependence            & long memory          & long memory          \\
		fractal dimension     & $\begin{cases} 2-H_1,& t< \tau\\ 2-H_2, & t >\tau\end{cases}$ & $\begin{cases} 2-H_1,& t< \tau\\ 2-H_2, & t >\tau\end{cases}$ \\
		MSD                   & $\begin{cases} Dt^{2H_1},& t\le \tau\\ Dt^{2H_2}, & t >\tau\end{cases}$                   &    $\begin{cases} D_1t^{2H_1},& t\le \tau\\  \sim D_1\tau^{2H_1}+D_*(t-\tau)^{2H_*}, & t >\tau\end{cases}$              \\
		trajectories          & discontinuous at $\tau$        & continuous          
	\end{tabular}
\caption{Comparison of typical already established MFBM models and IMFBM for one change of Hurst exponent at time $\tau$. In the line for MSD $H_*\equiv \min\{H_2,1/2,(H_1+H_2)/2 \}$ and constant $D_*$ can be read from the full formula \refeq{stepMSD}. Note that for different variants of MFBM the exact behaviour in the left column my differ, the features given are typical, see\cite{Stoev2006}.}\label{table}
\end{table}

IMFBM is a generalisation of both FBM and scaled Brownian motion used to
model diffusion with changing diffusivity. The transitions of $H$ and $D$
can be both smooth or discontinuous. It is Gaussian. The associated MSD
and covariance can always be expressed as integrals which for the crucial
case of step step function protocols for $H$ and $D$ reduce to elementary
functions. The local fractal dimension of the IMFBM trajectories is $2-\mH_t$.

Mathematical models such as IMFBM are indispensable in creating objective
tools to determine the best combination of stochastic models and their
parameters given measured data. They are becoming  increasingly important
with the fast growing numbers of increasingly refined experiments in
complex systems. Some of the existing solutions are provided, e.g., by
Bayesian analyses \cite{samu,kevin} or the machine learning apparatus
\cite{henrik,gorka,andi1,janusz}. We provide a guide towards statistical
estimation of the transport parameters in our IMFBM model in the Supplementary
Material.

The IMFBM process can be used to model the data directly but we also see
a potential value in using it to construct more sophisticated tools. In
many systems the evolution of $H$ and $D$ should be considered to be random
itself. This is an example of a doubly stochastic modelling approach which
is straightforward to introduce into anomalous diffusion studies using
IMFBM. Another classical and indispensable tool in stochastic modelling are
Langevin equations which are a class of stochastic differential equations
suited to describe the diffusion phenomena. It seems very natural to use
IMFBM as a noise in those equations which would then allow us to study the
stochastic motion in the presence of an external potential and which fulfils
local fluctuation-dissipation relations.

\ack
RM was supported by the German Ministry for Education and Research (BMBF, grant
STAXS).

\section*{References}
\bibliographystyle{iopart-num}
\bibliography{bib}

\clearpage

\setcounter{page}{1}
\setcounter{section}{0}
\setcounter{equation}{0}
\setcounter{figure}{0}
\setcounter{table}{0}
\renewcommand{\theequation}{S\arabic{equation}}
\renewcommand{\thefigure}{S\arabic{figure}}
\renewcommand{\thetable}{S\arabic{table}}

\thispagestyle{empty}

\noindent
{\Large\textbf{Minimal model of diffusion with time changing Hurst\\[0.2cm]
exponent}}\\[0.2cm]
{\large\textbf{Supplementary material}}\\[0.64cm]
\hspace*{3.2cm}\textbf{Jakub {\'S}l\k{e}zak$^\dagger$ and Ralf
Metzler$^{\ddagger,\sharp}$}\\[0.12cm]
\hspace*{3.2cm}\footnotesize{
$\dagger$ Hugo Steinhaus Center, Wroc{\l}aw University of Science and
Technology, Poland\\
\hspace*{3.2cm}$\ddagger$ Institute of Physics \& Astronomy, University of
Potsdam, Germany\\
\hspace*{3.2cm}$\sharp$ Asia Pacific Centre for Theoretical Physics, Pohang
37673, Republic of Korea}\\

\section{Simulation}

We divide simulation methods of the IMFBM into two cases: the situation when
the covariance of $B_\mH$ can be efficiently calculated and when it cannot.
Let us start from the former case.

The stochastic integral (6) which represents $\dd B_\mH$ and
its covariance (3) are singular because they are defined in the
limit of infinitesimal $\dd t$. This singularity can be removed if one takes
one step back and approximates this process at small but fixed $\Delta t$.
Repeating the derivation of IMFBM, finite differences of FBM from equation
(2) lead to the numerator $\e^{\I\omega(t+\Delta t)}-\e^{\I
\omega t}$ instead of the limiting form $\I\omega\e^{\I\omega t}$.
Substituting $D\to\mD_t,H\to\mH_t$ yields
\begin{equation}
\label{aIMFBMdef}
\Delta\widetilde B_\mH(t)\equiv\frac{\sqrt{\mD_t}}{\gamma_{\mH_t}}\int\limits
_{-\infty}^\infty\frac{\e^{\I\omega \Delta t}-1}{|\omega|^{\mH_t+1/2}}\e^{\I
\omega t}\dd Z(\omega).
\end{equation}
For this process the $\propto \omega$-dependence of the numerator at $\omega
\to0$ cancels the singularity of the denominator and allows for a stable
simulation. Given this process, it can then be used to generate $\Delta
\widetilde B_\mH(t)$. Then one can calculate $\widetilde B_\mH(n\Delta
t)\equiv\sum_{k=1}^n\Delta\widetilde B_\mH(k\Delta t)$. This process
approximates $B_\mH$ as $\Delta t\to0$.

The integral \refeq{aIMFBMdef} can be approximated directly by Riemann
summation over a finite \XXX{sequence of iid $\mathcal N(0,\Delta \omega)$
variables } $\Delta Z(\omega)$, preferably probing different
$\omega$ at logarithmic spans in order to better catch the power-law
distributed probabilistic mass. Alternatively, the process $\Delta
\widetilde B_\mH$ has the covariance
\begin{equation}
\langle\Delta\widetilde B_\mH(s)\Delta\widetilde B_\mH(t)\rangle=
\frac{1}{2}c_{\mH_s,\mH_t}\sqrt{D_sD_t}\cdot\lt(|t-s+\Delta t|^{\mH_s
+\mH_t}+|t-s-\Delta t|^{\mH_s+\mH_t}-2|t-s|^{\mH_s+\mH_t}\rt).
\end{equation}
Given this formula, its realisations can be simulated exactly using the
Cholesky decomposition method \cite{Kroese}.

For a given interval $a\le t\le b$ containing $n$ $\Delta t$-spaced points
we first need to store the covariance matrix $\Sigma_{i,j}\equiv\langle\Delta
\widetilde B_\mH(a+i\Delta t)\Delta\widetilde B_\mH(a+j\Delta t)\rangle$
which takes $\mathcal O(n^2)$ memory space. Using one of the standard Cholesky
decomposition algorithms we calculate the lower triangular matrix $L$ such that
$\Sigma=LL^\dagger$. These algorithms use $\mathcal O(n^3)$ time. Finally,
applying this matrix to the vector of iid $\mathcal N(0,1)$ variables returns
one realisation of $\Delta \widetilde B_\mH$. This method is initially memory
and time consuming with respect to $n$ but then generating new realisations
is very fast. One also needs to be careful simulating samples with low local
$\mH_t$, around $0.1$, because the dependence between increments is then
chaotic and for longer times may be dominated by the numerical errors of
the Cholesky decomposition, leading to false normal diffusion. This regime
of very low Hurst exponents is however rarely met in practice.

In the former category of simulations, when the covariance of $B_\mH$ can be
effectively determined, it is better to generate samples of $B_\mH$ directly,
without $\Delta\widetilde B_\mH(t)$. This can be done again by the Cholesky
decomposition and for any choice of sampling times $t_1,t_2,\ldots,t_n$, not
necessarily equally spaced. The covariance of $B_\mH$ can be numerically
approximated using an integral of type (4), but one then needs to
be careful, because, as a result of errors, the resulting matrix $\Sigma$
may not be exactly positive semidefinite, a requirement for the Cholesky
decomposition. Additional regularisation may be required beforehand.

\XXX{A perfect situation occurs when the covariance integral (4) can be calculated analytically.} This is the case for any stepwise protocol of constant $(H_j,D_j)$
on intervals $(\tau_{j-1},\tau_j)$. Without any loss of generality we can
assume that $s=\tau_n$ and $s=\tau_m$ (by taking $H_{n-1}=H_n,D_{n-1}=D_n$)
which allows us to formulate the result in an elegant manner,
\begin{multline}
\label{fullCov}
\langle B_\mH(s)B_\mH(t)\rangle=\\
\sum_{j=1}^{n}\sum_{k=1}^{m}A_{j,k}\big(|\tau_j-\tau_{k-1}|^{H_j+H_k}+|\tau_{
j-1}-\tau_{k}|^{H_j+H_k}-|\tau_{j-1}-\tau_{k-1}|^{H_j+H_k}-|\tau_{j}-\tau_{k}
|^{H_j+H_k}\big),  
\end{multline}
where we denoted $A_{j,k}\equiv c_{H_j,H_k}\sqrt{D_jD_k}/2$. In particular for
the case of only a single change of parameters at time $\tau$ the above formula
produces
\begin{multline}
\langle B_\mH(s)B_\mH(t)\rangle=\\
\frac{1}{2}\begin{cases}D_1\big(t^{2H_1}+s^{2H_2}-|t-s|^{2H_1}\big), & s\le t
\le\tau\\D_1\big(s^{2H_1}+\tau^{2H_1}-|\tau-s|^{2H_1}\big)+\\
\ c_{H_1,H_2}\sqrt{D_1D_2}\big(|\tau-s|^{H_1+H_2}+t^{H_1+H_2}-\tau^{H_1+H_2}-
|t-s|^{H_1+H_2}\big), & s<\tau<t\\
2D_2\tau^{2H_1}+\\
c_{H_1,H_2}\sqrt{D_1D_2}\big(s^{H_1+H_2}+t^{H_1+H_2}-2\tau^{H_1+H_2}-|t-\tau|^{
H_1+H_2}-|s-\tau|^{H_1+H_2}\big)+\\
D_2\big(|s-\tau|^{2H_2}+|t-\tau|^{2H_2}-|t-s|^{2H_2}\big), & \tau\le s\le t.
\end{cases}
\end{multline}
One last point we have to make about the numerical approach to IMFBM is
that the memoryless case $(\mH_s+\mH_t)/2=1/2$ is singular in the same way as
for simulations of classical FBM. Most of the methods which explicitly use the
kernel $|t-s|^{\mH_s+\mH_t-2}$ will result in numerical errors. One needs to
consider only "perturbed memorylessness" $(\mH_s+\mH_t)/2=1/2+\epsilon$, or,
preferably whenever possible, implement this case separately, by simulating
independent increments \cite{FBMsim}.

A computer code for IMFBM simulation is provided in in the GitHub repository
{\url{https://github.com/jaksle/IMFBM}}.

\section{Estimation}

We provide two estimation methods for IMFBM which have different
requirements and complement each other. They are designed for experimental
data measured at constant rate, $X_j=B_\mH(\Delta tj),j\in\{0,1,2,\ldots,N\}$.
We denote the corresponding increments by $\Delta X_j\equiv X_{j+1}-X_j$.

In the first case we discuss a method which is of use even if only one
trajectory is available and which can detect variability of subdiffusive/normal
$H,D$. For stepwise changes it estimates their local values and transition
moments. It is based on the second variation and lag 1 covariation of the
time series,
\begin{equation}
\mV_k\equiv\sum_{i=0}^k(\Delta X_i)^2,\qquad\mC_j\equiv \sum_{i=0}^k\Delta X_i
\Delta X_{i+1}.
\end{equation}
Due to the law of large numbers, for big $k$, $\mV_k$ is asymptotically
dominated by a straight line with slope equal to the variance of the series
increments $\mV_k=\sigma^2k+\mathcal{O}(\sqrt k),\sigma^2\equiv\langle(\Delta
X_i)^2\rangle$, similarly with a covariance with slope equal to the lag 1
covariance which we separate into variance and correlation factors $\mC_k=
\sigma^2\rho k+\mathcal{O}(\sqrt k),\rho\equiv\langle\Delta X_i\Delta X_{i+1}
\rangle/\langle(\Delta X_i)^2\rangle$. Thus, fitting a linear function to these
quantities yields the estimates $\sigma^2_\text{est}$ and $\rho_\text{est}$.

In the final step, the Hurst exponent is reconstructed from the correlation
$\rho_\text{est}$. The covariance formula (3) is valid for
infinitely small increments, for finite sampling frequency a correction should
be included, resulting in the relation $\rho=2^{2H}-2$ and $H_\text{est}
=\log_2(2\rho_\text{est}+2)/2$. Given the estimate $H_\text{est}$ the
diffusivity is $D_\text{est}=\sigma^2_\text{est}/(\Delta t)^{2H_\text{est}}$.

At transition moments of $H$ or $D$, the slopes of $\mV$ and $\mC$ change
as well and they become polygonal chains; \XXX{this is caused by the fact
that IMFBM is locally undistinguishable from FBM, and slope is a local
property}. The estimation procedure is then
as follows: plot $\mV,\mC$ and asses if their slope changes; this suggests
an IMFBM transition. Fit a polygonal chain to $\mV$ and $\mC$ using, e.g.,
the ordinary least squares method. Read the transition moments and evaluate
estimates of the local Hurst exponent and the diffusivity from the slopes of the
chain segments.

\begin{figure}
\centering
\includegraphics[width=10cm]{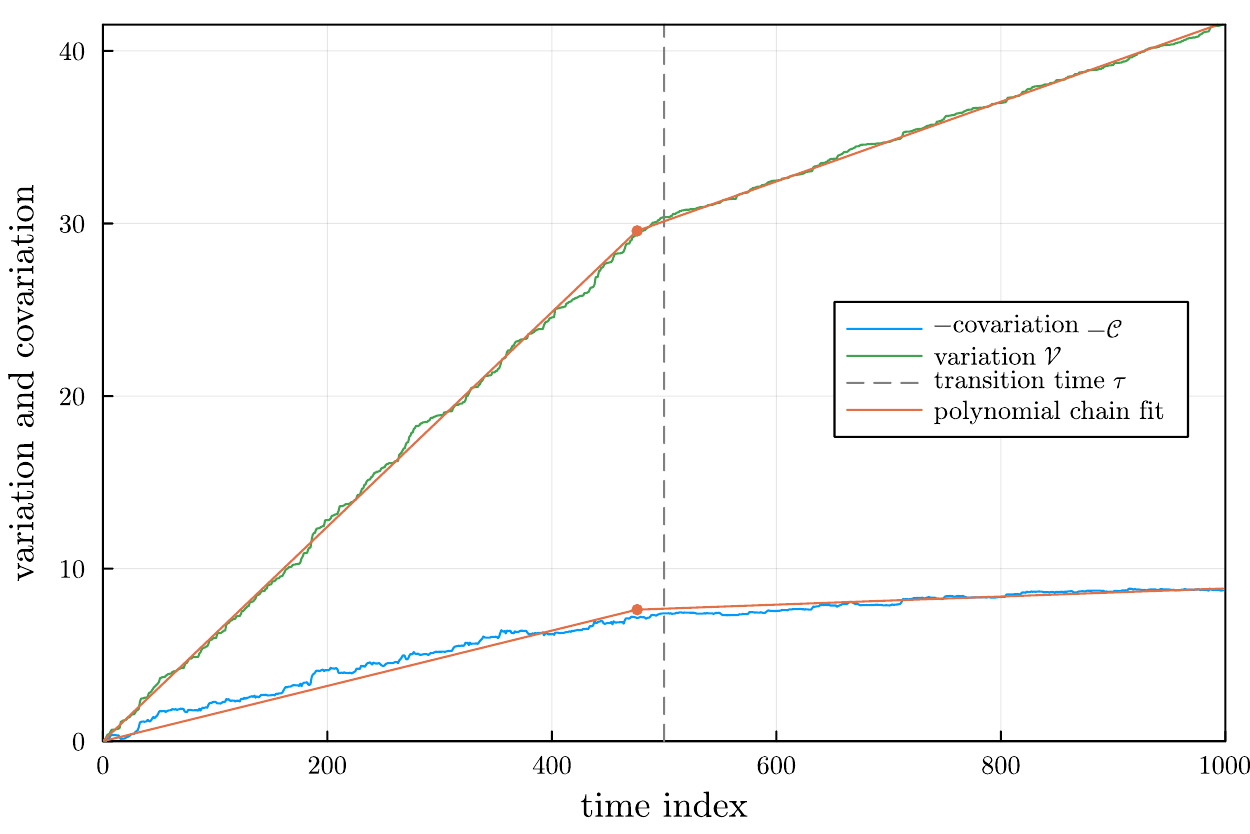}
\caption{Estimation for a single trajectory of length 1000. For better visual
presentation minus the covariation is shown. The obtained estimate for the
transition time is $\tau_\text{est}$=476, comparing nicely with the true value
500. From the slopes of the fitted polynomial chain we obtained estimates of
the Hurst exponents and diffusivities $H_{1,\text{est}}=0.28,D_{1,\text{est}}
=0.86,H_{2,\text{est}}=0.42,D_{2,\text{est}}=1.13$.}
\label{polyFit}
\end{figure}

We tested this method using trajectories of IMFBM at $0<t<10$, which change
from $H_1=0.3,D_1=1$ to $H_2=0.45,D_2=1.5$ in the middle. One exemplary
estimation is visualised in figure \ref{polyFit}. Using a sample of 1000
realisations of IMFBM we measured the behaviour of the estimators for different
lengths of trajectories. The results are listed in Table \ref{estTable}
which proves that this method allows for reliable estimation using a very
moderate amount of data.

\begin{table}
\centering
\begin{tabular}{l|c|c|c|c|c}
&$H_1$ & $D_1$ & $\tau$ & $H_2$ & $ D_2$ \\ \hline
$n=500$ &$0.29\pm 0.06$ & $ 1.0\pm 0.5$ & $244\pm 30$ & $0.44\pm 0.06$ & $1.6\pm
0.7 $\\
$n=1000$ &$0.29\pm 0.04$ & $ 1.0\pm 0.4$ & $494\pm 32$ & $0.44\pm 0.05$ & $1.6\pm
0.7 $\\
$n= 10 000$ &$0.30\pm 0.01$ & $1.0\pm 0.2$ & $5001\pm 63$ & $0.45\pm0.02$ &
$1.5\pm0.5 $\\
\end{tabular}
\caption{Means and mean squared deviations of the IMFBM estimators for
trajectories with different lengths $n$. The switching times were chosen
at the mid-point of each trajectory.}
\label{estTable}
\end{table}

The limitations of this method are twofold. First, it does not work for
superdiffusion for which due to the long memory variation and covariation are
not linear. Second, it is sensitive to high frequency data distortions due to
using only a lag 1 correlation. For example, additive noise would distort the
estimation of the Hurst exponent by weakening the short range correlation;
however, the transition moment would still be correctly detected which then
opens the possibility to use classical statistical methods for FBM at each
detected segment. The method can also be generalised to use correlations at
different lags.

The second method uses a sample of trajectories which share $\mH$ and $\mD$,
or at least we want to estimate some form of effective, averaged $\mH$ and
$\mD$ from the sample. Conceptually this method is very straightforward:
given a sample we estimate the local MSD, fit it with a power law and from
this obtain local estimates of the Hurst exponent and diffusivity.

\begin{figure}\centering
\includegraphics[width=8cm]{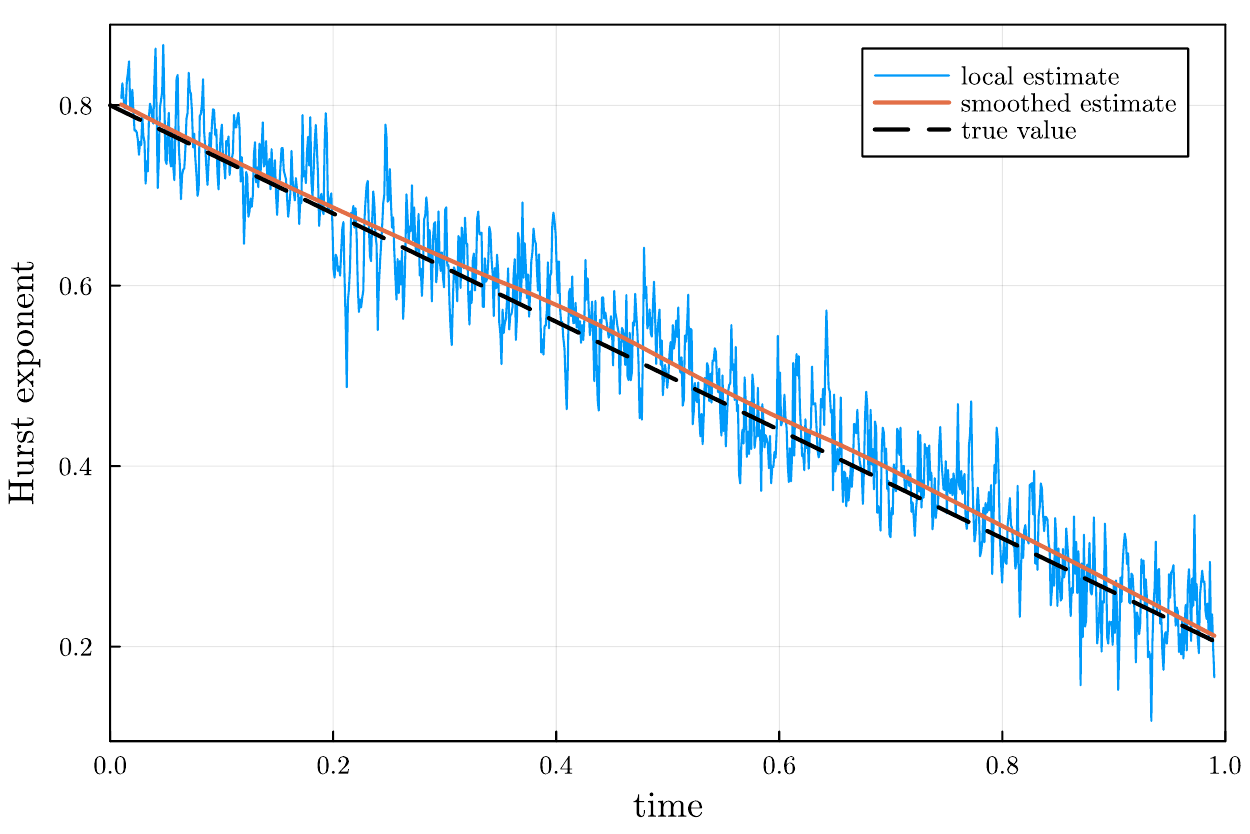}
\includegraphics[width=8cm]{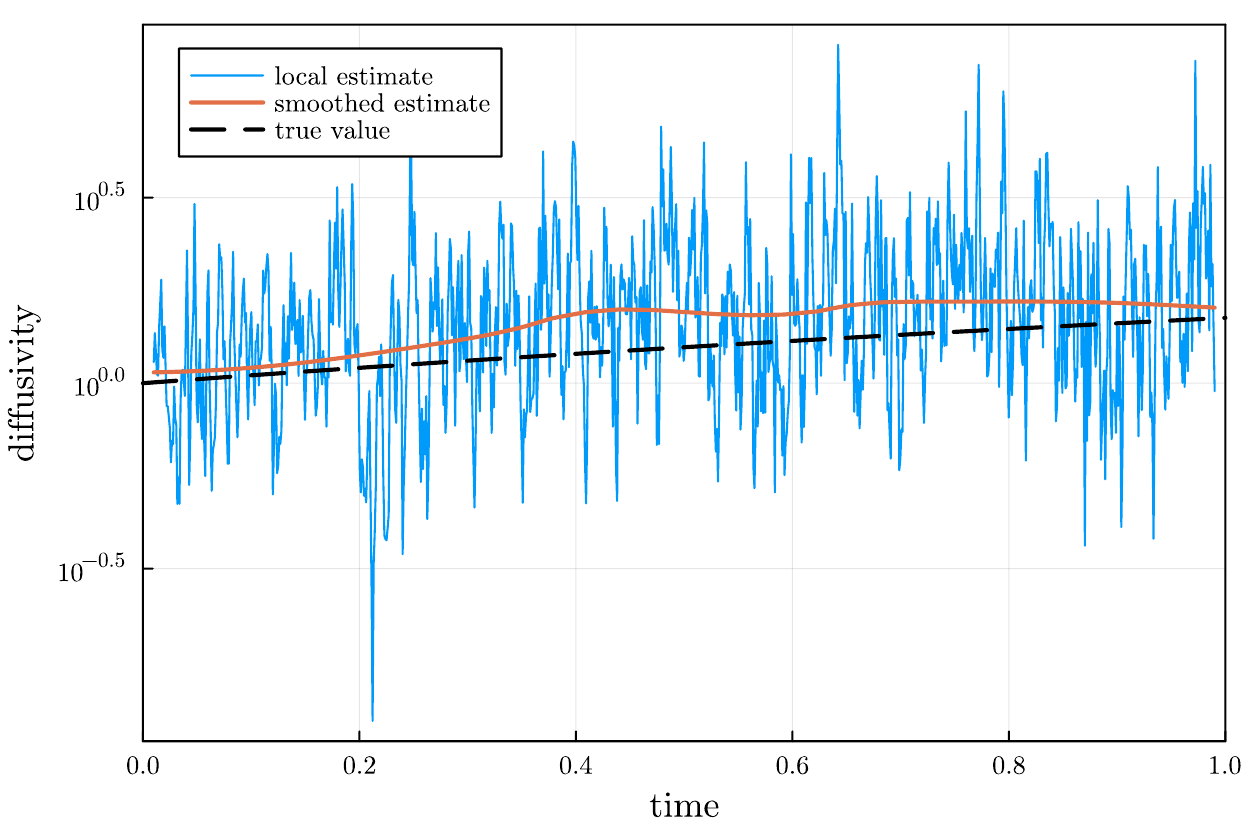}
\caption{Estimation based on a sample of 100 trajectories with lengths 1000
each. The Hurst index was decreasing linearly from 0.8 to 0.2, the diffusivity
was increasing linearly from 1 to 1.5. The estimation window was $w=10$.
A smoothed estimate was obtained using the Loess method. Because of the
multiplicative nature of the $D$ estimation errors smoothing and data
visualisation were performed in the $\log y$ scale. Mean squared deviations of
local estimates were $\pm0.04$ for $H$ and $\pm0.5$ for $\log D$. Mean errors
were $0.01$ and $0.15$, respectively.}
\label{locEst}
\end{figure}

By "local MSD" we mean the quantity $\delta_{s,t}^2\equiv\langle(X_{s+t}-
X_s)^2\rangle$ which is also considered in literature under the names
"structure function" or "semivariogram". For IMFBM, if $H$ and $D$ are
constant in the interval $(s,s+t)$ and equal to $\mH_s,\mD_s$ the local MSD
reads $\delta_{s,t}^2=\mD_st^{2\mH_s}$. It can be estimated as sample average
$\overline{(X_{j+k}-X_j)^2}$ and fitted as function of $k\in\{0,1,2,\ldots,w\}$
locally in a window $w$. The reliability of the estimation requires that $\mH,
\mD$ do not
vary too much over this window. Choosing a proper $w$ requires balancing
the expected variability of the parameters and the MSD power law fit quality. We
show an exemplary estimation in figure \ref{locEst}.

Computer code examples of IMFBM estimation are provided in the GitHub repository
{\url{https://github.com/jaksle/IMFBM}}.

\end{document}